\documentclass[prl,showpacs,showkeys,amsmath,amssymb,twocolumn]{revtex4}

\usepackage{dcolumn}
\usepackage{bm}
\input xy
\usepackage[all]{xy}
\xyoption{2cell} \UseAllTwocells

\begin{document}



\title{Index theorem, spin Chern Simons theory and fractional magnetoelectric effect
\\in strongly correlated topological insulators}
\author{K.-S. Park}
\email{kpark@postech.ac.kr}
\author{H. Han}
\affiliation{Department of Electrical Engineering, Pohang University
of Science and Technology, San 31, Hyojadong, Namgu, Pohang 790-784,
Korea}

\date{\today}

\begin{abstract}
Making use of index theorem and spin Chern Simons theory, we
construct an effective topological field theory of strongly
correlated topological insulators coupling to a nonabelian gauge
field $ SU(N) $ with an interaction constant $ g $ in the absence of
the time-reversal symmetry breaking. If $ N $ and $ g $ allow us to
define a t'Hooft parameter $ \lambda $ of effective coupling
as $ \lambda = N g^{2} $, then our construction leads to the
fractional quantum Hall effect on
the surface with Hall conductance $ \sigma_{H}^{s} =
\frac{1}{4\lambda} \frac{e^{2}}{h} $.
For the magnetoelectric response described by a bulk axion angle $ \theta $,
we propose that the fractional magnetoelectric effect can be realized in
gapped time reversal invariant topological insulators of strongly correlated
bosons or fermions with an effective axion angle
$ \theta_{eff} = \frac{\pi}{2 \lambda} $ if they can have
fractional excitations and degenerate ground states on topologically nontrivial
and oriented spaces. Provided that
an effective charge is given by $ e_{eff} = \frac{e}{\sqrt{2 \lambda}} $,
it is shown that $ \sigma_{H}^{s} = \frac{e_{eff}^{2}}{2h} $,
resulting in a surface Hall conductance of gapless fermions with $ e_{eff} $
and a pure axion angle $ \theta = \pi $.
\end{abstract}


\maketitle

Recently topological insulators (TIs) have received a great deal of
attention \cite{moore1,kane1,qz}. The theory of TI has been
developed along a few different directions. As one of developments,
topological band theory has been described in terms of $ Z_{2} $
topological invariants for noninteracting band insulators in (3 + 1)
dimensions (Ds) when the time-reversal symmetry (TRS) is not broken
\cite{kane2,kane3,moore2,roy}. TI theory has also been proposed in
HgTe quantum wells \cite{bernevig} in the presence of spin-orbit
interactions. Furthermore, there have been many experimental
observations supporting the existence of nontrivial topological
surface states in numerous materials
\cite{moore1,kane1,qz,konig,chen}.

As another direction of TI theory, topological field theories (TFTs)
have been suggested in the low energy limit of TIs \cite{zhang,qi1}.
In particular, analogous to the coupling of an axion particle to
ordinary electric and magnetic fields, the partition function of TFT
is expressed by
\begin{eqnarray}
\mathcal{Z}(F) = C \mathrm{exp}(iK_{\theta} \frac{e^{2}}{4 \pi}
\int_{M_{4}} F \wedge F ),
\end{eqnarray}
where $ C $ is a constant, and $ F $ is the electromagnetic field
strength \cite{wilczek2,qi1}. Here $ K_{\theta} $ is a bulk
magnetoelectric polarization  written by $ K_{\theta} =
\frac{\theta}{2 \pi}$. Under shifts of $ \theta $ by multiples of $
2 \pi $, the partition function and all physical quantities are
invariant on the periodic boundary condition. But the $ \theta $
term has a problem dangerous to the TRS. It follows that trivial
insulators have $ \theta = 0 $ mod $ 2 \pi $ whereas noninteracting
TIs take $ \pi $ mod $ 2 \pi $ in the values allowed by a time (T)
operation. Without breaking the TRS, the current theoretical
challenge is how to extend noninteracting TIs to the strongly
correlated electron systems of TIs \cite{qi1,essin,qi2}. It has more
recently been proposed that there is a possibility of T-invariant
fractional TIs for fermions and bosons in correlated systems
\cite{maciejko,swingle,park2}. More interesting issue is to
formulate the general TI theory for strongly correlated systems that
cannot be smoothly connected to any band insulator.

In this article, using flux quantization, index theorem and spin
Chern Simons (CS) theory, we present a fractional magnetoelectric effect
of strongly correlated TIs coupling to a non-abelian
gauge field $ SU(N) $ with an interaction constant $ g $ on an easy
and simple effective field theory (FT) under the TRS and
topological gauge invariance. The low energy FT can
provide the interacting $ SU(N) $ gauge
theory in TIs unless the TRS is spontaneously broken. If a t'Hooft
parameter $ \lambda = Ng^{2} $ is defined
as an effective interaction constant, then our construction can lead to the
fractional quantum Hall effect (FQHE) on the surface with Hall
conductance $ \sigma_{H}^{s} = \frac{1}{4\lambda} \frac{e^{2}}{h}$.
In the viewpoint of the magnetoelectric
response for the interacting TIs, the fractional axion angle allows
us to result in the fractional magnetoelectric polarization $
K_{\theta} = \frac{1}{4 \lambda} $. It is proposed that the
factional magnetoelectric effect can be realized in gapped time
reversal invariant topological insulators of strongly correlated
bosons or fermions with an effective axion angle
$ \theta_{eff} = \frac{\pi}{2\lambda}$ if they can
have fractional excitations and degenerate ground states on
topologically nontrivial and oriented spaces.

On the other hand, provided that an effective charge can be expressed in terms of
$ e_{eff} = \frac{e}{\sqrt{2 \lambda}} $, the surface Hall
conductance is regarded as a conductivity $ \sigma_{H}^{s} =
\frac{e_{eff}^{2}}{2h} $ for a single Dirac cone of gapless fermions
with fractional charge $ e_{eff} $ and a pure axion angle $ \theta = \pi $.
Recently, angular photoemission spectroscopy experiment has showed that
there is a topological phase transition from a trivial insulator into
a topological surface state by displaying the emergence of
spin vortex with fractional charge $ \pm \frac{e}{2} $  and $ \theta = \pi $
in a tunable topological insulator $ \mathrm{BiTl(S_{1-\delta}Se_{\delta})} $ \cite{xu}.
Theoretically it has been suggested that there can be a quantized vortex of fractional charge
$ \pm \frac{e}{2} $ and an odd number of gapless Dirac fermions at the surface
of a strong TI \cite{seradjeh}. As a special viewpoint of these experimental
and theoretical studies, when $ \lambda = 2 $, our theoretical construction shows
that the TIs can have fractionalized charge $ \frac{e}{2} $ topological
objects with a bulk gap and string-like vortex excitations.
These topological excitations can be described in terms
of a deconfined $ Z_{2} $ gauge theory in (3+1)D.

The article is organized as followings.
We will review parton models of effective FT for correlated TIs in Section II.
In Section III, Dirac quantization conditions will be illuminated
in the cases of complex spinor fields on a Riemannian manifold with
spin structures and nonempty spin boundary. For complex spinor fields on the spin surface,
the Chern-Simon (CS) theory and Atiyah-Patodi-Singe (APS) index theorem will be
exploited to investigate the effective QFT for the interacting nonabelian gauge fields
of correlated TIs in Section IV. In V, we will describe a fractional surface
Hall conductance of correlated TIs and will use the APS theorem
to explain a general form for the Hall conductance and magnetoelectric effects
on a Riemann spin surface with a genus in the effective QFT.
And finally we will come to summary and conclusion for our results in Section VI.

\section{\label{sec:level1} II. Review of Parton Models for Effective Field Theory}

Let us review parton models of effective FT in interacting topological insulators.
Then for a more systematical approach of the interacting TIs,
we take into account the projective construction of a correlated electron system
for building the effective FT on a Riemann surface emerged
from a (3 + 1)D Riemannian manifold with spin structures.
An electron is decomposed into $ N $ different fractionally charged
and fermionic partons \cite{wen,barkeshli}.
It is without loss of generality that the partons
generate a ground state of topological phases. When the partons get together
to create the physical real electrons, a new ground state of topological phase
can appear from their recombination.
Provided that the partons are recombined together to represent the
physical electrons, we can construct an interacting many-body
wavefunction as a new topological state of electrons emerged from (3 +
1)D.

One can generalize this construction to $ N_{f} $
different flavors of charged fermion partons, with $
N_{f}^{c} $ partons for each flavors $ f = 1, \cdots, N_{f} $.
The electron is fractionalized into $ N_{f} $
different flavors of fractionally charged fermion partons, with $
N_{f}^{c} $ partons for each flavors. Under
these decompositions, we should obey two crucial constraints. First,
because the electron preserves the fermion statistics, the
total number of partons per electron has to be odd such that
\begin{eqnarray}
 N_{1}^{c} + \cdots + N_{N_{f}}^{c}  = \mathrm{odd}.
\end{eqnarray}
The second constraint is that the total charge of the partons should
sum up to the electron charge $ e $ such as
\begin{eqnarray}
N_{1}^{c} q_{1} + \cdots + N_{N_{f}}^{c}q_{N_{f}} = e
\end{eqnarray}
when $ q_{f} < e $ is the fractional charge for partons of flavor $
f $.
The total electron wavefunction is expressed by a product of
parton ground state wavefunctions \cite{jain}
\begin{eqnarray}
\prod_{f=1}^{N_{f}} \Psi_{N_{f}^{c}}(
\{\mathrm{r}_{i},\mathrm{s}_{i}\} ) =\prod_{f=1}^{N_{f}} \lbrack
\Psi_{f}( \{ \mathrm{r}_{i},\mathrm{s}_{i} \} ) \rbrack^{N_{f}^{c}}.
\end{eqnarray}
Here $ \Psi_{f}( \{\mathrm{r}_{i},\mathrm{s}_{i}\} ) $ stands for
the parton ground state wavefunction given by a Slater determinant
which describe the ground state of a noninteracting TI Hamiltonian,
and $ \{\mathrm{r}_{i},\mathrm{s}_{i}\}, i = 1, \cdots, N $, the
position and spin coordinates of the partons.

For a more specific approach for strongly correlated TI on a lattice
of $ SU(N) $ electrons, the Hamiltonian is expressed by
\begin{eqnarray}
H = \sum_{ij} \{ C^{\dagger}_{i \alpha} h^{\alpha \beta}_{ij}
e^{ieA_{ij}} C_{j \beta} + H. C. \} + H_{int}(C^{\dagger},C),
\end{eqnarray}
where $ i, j $ stands for site indices, $ \alpha, \beta $ denotes
internal degrees of freedom, $ h_{ij} $ indicates the Hamiltonian
matrix, $ A_{ij} = \int_{\mathrm{r}_{i}}^{\mathrm{r}_{j}}
d\mathrm{r} \cdot \mathrm{A} $ with the $ U(1) $ electromagnetic
vector potential $ \mathrm{A} $. $ H_{int} $ denotes an interaction
Hamiltonian between electrons. $ C_{i \alpha} $ is the electron
operator that is decomposed into \cite{jain}
\begin{eqnarray}
C_{i \alpha} = \prod_{f=1}^{N_{f}} \psi^{f}_{1
\alpha}(\mathrm{r}_{i}) \cdots \psi^{f}_{ N_{f}^{c} \alpha
}(\mathrm{r}_{i}) \end{eqnarray} when satisfying the constraint
rules of Eqs. (2) and (3).

It is well known that the quark operators can act on a Hilbert space
larger than the physical electron one. The unphysical states which
cannot become invariant under unitary transformations, should
be removed from the quark Hilbert space. When the electron
operators become preserved, those transformations provide the
$ SU(N_{f}^{c}) $ for quarks in the $ N_{f}^{c} $ representation
of each flavor $ f=1, \cdots, N_{f} $. Thus it follows that the projection
can be taken onto the electron Hilbert space implemented by
coupling minimally the quark to a $ SU(N_{f}^{c}) $ gauge field $
a_{\mu} $ with an interacting constant $ g $.
This gives rise to observation only of $ SU(N_{f}^{c}) $ excitations in its
low-energy spectrum. Quarks of a given flavor can be symmetric or
antisymmetric in their $ N_{f}^{c} $ odd color indices
with the constraints of Eq. (2) for $ f=1, \cdots, N_{f} $.

\section{\label{sec:level1} III. Dirac Quantization Condition of Spinor Fields}

Now we begin with the simplest case of $ N_{f} = 1 $ and $ N^{c}_{1} $ odd.
Dirac quantization of flux can have a nice way of topological gauge invariance
in the representation of complex spinor fields of
antisymmetric $ N_{1}^{c} $ partons with $ N^{c}_{1} $ odd
for a composite electron in the interacting TI.
Let us consider the Dirac quantization of fermions
represented by complex spinor fields through a two-cycle $ \Sigma $ in
the sense of antisymmetric $ N^{c}_{1} $ partons for a composite electron with the
odd-number constraint of $ N^{c}_{1} $ on $ M_{4} $.
Then it follows that we don't
necessarily require a real (neutral) spinor field on $ M_{4} $. The spinor fields
can have a natural connection to a possible obstruction of spin which can be described in
terms of the second Stiefel-Whitney class $ \omega $ as an
element of $ \mathbb{\check{H}}^{2}(M_{4},\mathrm{Z}_{2}) $ which
is called the second cohomology group with a coefficient $ \mathrm{Z}_{2} $
over $ M_{4} $. In order to exist neutral spinors on $ M_{4} $,
it is shown that its vanishing, i.e., mod 2, is required
as a necessary and sufficient condition for them.
On the condition that $ \omega $ is well defined to mod 2,
$ M_{4} $ has a $ spin $ structure.

Assume that there are complex spinor fields on $ M_{4} $. Then they
provide a deep insight for Dirac's quantization condition.
Under the Dirac quantization of flux, the antisymmetric parton wavefunctions
can be represented by $ SU(N_{1}^{c}) $ spinors on $ M_{4} $ which
is covered by a finite number of neighborhoods $ U_{i} $ for $ i = 1, 2, \dots, L $.
In each neighborhood, more structures have to be taken into account on the
representation of internal symmetries. For more structure, in addition to the $
U(1) $ connection or gauge potential, we should consider an oriented
frame of vierbein $ V_{i} $ for $ i = 1, 2, \dots, L $, and complex spinor fields of
antisymmetric partons $ \{ \Psi_{1i} \}^{N_{1}^{c}} $ with $
N_{1}^{c} $ only odd. These symmetries can be dependent on
choices made in the neighborhood $ U_{i}$. As choices of degrees of
freedom, there can exist local $ U(1)$ gauge transformations $
\mathit{\chi} $
\begin{eqnarray}
\{ \Psi_{1i} \}^{N_{1}^{c}} \rightarrow \{ e^{iq_{1} \mathit{\chi_{i}}}
\Psi_{1i} \}^{N_{1}^{c}}, \quad A_{i} \rightarrow A_{i} + d
\mathit{\chi_{i}},
\end{eqnarray}
for $ i = 1, 2, \dots, L $, and $ SU(N_{1}^{c}) $
gauge transformations $ \mathit{\lambda} $
\begin{eqnarray}
V_{i} \rightarrow \mathrm{R} V_{i}, \quad
a_{i} \rightarrow a_{i} + d \mathit{\lambda_{i}}, \nonumber\\
\{ \Psi_{1i} \}^{N_{1}^{c}} \rightarrow \{ \mathrm{S}( \mathrm{R} )
e^{ ig \mathit{\lambda_{i}} }  \Psi_{1i} \}^{N_{1}^{c}},
\end{eqnarray}
where $ \mathrm{R}  \in SO(4) $ and for $ i = 1, 2, \dots, L $.
Here $ q_{1} $ and $ g $ have indicated an electric charge of one flavor
and an interaction constant of $ SU(N_{1}^{c}) $, respectively.  It is easily
seen that there can be a sign ambiguity from the lift of $ \mathrm{R}
\rightarrow \pm \mathrm{S}( \mathrm{R} ) $ since the quotient of the
spin group $ Spin(4)$ by $ \mathrm{Z}_{2} $ is isomorphic to
$ SO(4) $. In the sense of a double overlap on two
contiguous neighborhoods, $ U_{i} \cap U_{j} \ne 0 $, one must
take transition functions associated with transformation groups
\begin{eqnarray}
A_{i} \rightarrow A_{j} + d \mathit{\chi}_{ij}, \quad a_{i}
\rightarrow a_{j} + d \mathit{\lambda}_{ij}, \quad
V_{i} \rightarrow \mathrm{R}_{ij} V_{j}, \nonumber \\
\{ \Psi_{1i} \}^{N_{1}^{c}} \rightarrow \{ \mathrm{S}(
\mathrm{R}_{ij} )e^{iq_{1} \mathit{\chi}_{ij}} e^{ig
\mathit{\lambda}_{ij}} \Psi_{1j} \}^{N_{1}^{c}},
\end{eqnarray}
for $ i = 1, 2, \dots, L $. There is without loss of generality to assume that
\begin{eqnarray}
\mathrm{R}_{ij} &=& ( \mathrm{R}_{ji} )^{-1}, \mathit{\chi}_{ij} =
(\mathit{\chi}_{ji})^{-1}, \nonumber\\
\mathit{\lambda}_{ij} &=& (\mathit{\lambda}_{ji})^{-1}, \mathrm{S}(
\mathrm{R}_{ij} ) = (\mathrm{S}( \mathrm{R}_{ji} ))^{-1},
\end{eqnarray}
for $ i, j = 1, 2, \dots, L $.
On a triple overlap region, $ U_{i} \cap U_{j} \cap U_{k} \ne 0,
\forall i, j, k = 1, 2, \dots, L $,
which is supposed to be contractible, one can have consistency
conditions
\begin{eqnarray}
 \mathrm{R}_{ij}\mathrm{R}_{jk}\mathrm{R}_{ki} = I,
 S(ijk) \equiv \mathrm{S}( \mathrm{R}_{ij} ) \mathrm{S}(
\mathrm{R}_{jk} ) \mathrm{S}( \mathrm{R}_{ki} ) = \pm I. \quad
 \end{eqnarray}
It follows that the above equations have identity elements of $
SO(4) $ and $ Spin(4)$.
This returns the frame $ V_{i} $ to itself.
Consequently, under $ U(1) \times SU(N_{1}^{c}) $ gauge
transformations and in the spinor representations of antisymmetric $
N^{c}_{1} $ partons with $ N^{c}_{1} $ odd, one can express
\begin{eqnarray}
\{ \Psi_{1i} \}^{N_{1}^{c}} &=& \{ e^{iq_{1} ( \mathit{\chi}_{ij} +
\mathit{\chi}_{jk} + \mathit{\chi}_{ki} )} e^{ig
(\mathit{\lambda}_{ij} + \mathit{\lambda}_{jk} +
\mathit{\lambda}_{ki} )} \nonumber\\
&& \mathrm{S}( \mathrm{R}_{ij} )
\mathrm{S}( \mathrm{R}_{jk} ) \mathrm{S}( \mathrm{R}_{ki} )
\Psi_{1i} \}^{N_{1}^{c}},
\end{eqnarray}
for $ i, j, k = 1, 2, \dots , L $.
In the above expression, the product of the three matrices $
\mathrm{S} $ can play an essential role, and so $ S(ijk) $ can indicate
the product as a following
\begin{eqnarray}
S(ijk) \equiv \mathrm{S}( \mathrm{R}_{ij} ) \mathrm{S}(
\mathrm{R}_{jk} ) \mathrm{S}( \mathrm{R}_{ki} ) = \pm I,
\end{eqnarray}
for $ i, j, k = 1, 2, \dots , L $. Up to a sign, a crucial point, is to take a lift from $
SO(4) $ to $ Spin(4) $ in the right hand side of Eq.
(13). One cannot decide the sign when it is dependent on the
choices made in the transformation groups of Eq. (13). In the sense
of different overlaps, the signs of Eq. (13) can not be totally
independent. Thus the spinor consistency condition leads one to
obtain
\begin{eqnarray}
e^{iq_{1} C_{ijk}} e^{ig D_{ijk}} = S(ijk),
\end{eqnarray}
in the representation of the complex antisymmetric wavefunctions with $ g $
in addition to the symmetric scalar ones with $ q_{1} $ in terms of $ C_{ijk} $.
In Eq. (14), $ D_{ijk} $ satisfies the self-consistency relation
\begin{eqnarray}
D_{ijk} \equiv \mathit{\lambda}_{ij} + \mathit{\lambda}_{jk} +
\mathit{\lambda}_{ki} \in  \frac{\mathrm{Z}}{g N_{1}^{c}}
\end{eqnarray}
for $ i, j, k = 1, 2, \dots , L $. Therefore, under the flux of $ U(1) \times
SU(N_{1}^{c}) $ gauge theory through a two-cycle $ \Sigma $, the
Dirac quantization of complex spinor fields produce
\begin{eqnarray}
\mathrm{exp}( 2\pi i \int_{\Sigma} ( q_{1} F + g G ) ) =
(-1)^{\omega(\Sigma)},
\end{eqnarray}
where $ G $ is the $ SU(N_{1}^{c}) $ field strength, and $ q_{1} =
\frac{e}{N_{1}^{c}} $ with $ N_{1}^{c} $ odd while $ F $ is the $ U(1) $.
Here the sign is determined by the finite product over triple overlaps
\begin{eqnarray}
(-1)^{\omega(\Sigma)} = \prod_{U_{i} \cap U_{j} \cap U_{k} \cap
\Sigma \ne 0} S(ijk).
\end{eqnarray}
Equation (16) is regarded as the preliminary expression for the Dirac
quantization condition of fermions in the complex spinor
representations of antisymmetric parton wavefunctions. The crucial idea
is how $ \omega(\Sigma) $ in question should be independent
of the choices made on the covering of two-cycle $ \Sigma $
by neighborhoods. Without loss of generality, the preliminary result
proves a natural connection for the Dirac quantization of flux certainly
independent of those choices.
Likely, $ \Sigma $ can not be changed under transformation of the
two-cycle by a homologous one $ \Sigma \rightarrow \Sigma^{\prime} +
\partial \Lambda.$ By all these results, it is shown that $
\Sigma $ can be closely associated with the Stiefel-Whitney two-cocycle
$ \omega $ over $ \mathrm{Z}_{2} $.

In general, let us extend a base $ \Sigma $ to a set of bases
$ \Sigma_{1}, \Sigma_{2}, \dots, \Sigma_{L} $ of
the integer lattice. Suppose that $
\omega_{i} = \omega(\Sigma_{i}), \forall i = 1, 2, \dots, L $.
Then the quantization condition is extended to get shifted into
\cite{alvarez,am,witten}
\begin{eqnarray}
\int_{\Sigma_{i}} (  \frac{e}{N_{1}^{c}} F + g G )- \frac{1}{2}
\omega_{i}, \forall i = 1, 2, \dots, L.
\end{eqnarray}
Therefore there can be possible on fractional values of flux
through a two-cocycle $ \Sigma $ in the total $ U(1) \times SU(N_{1}^{c}) $
gauge theory of quark fields described by antisymmetric $ N^{c}_{1} $ partons with $ N^{c}_{1} $
odd. One should still identify the number $ \omega_{i} $ with the
self-intersection one of $ \Sigma_{i} $ mod 2, $ \forall i = 1, 2, \dots, L $.

Finally we take into account the flux through $ \Sigma_{i} $ with
nontrivial boundary $ \partial \Sigma_{i}, \forall i = 1, 2, \dots,
L $. In the representation of complex spinor fields for electrons
formed by antisymmetric partons, the spinor field consistency Eq.
(18) gives rise to the boundary Dirac quantization condition
\begin{eqnarray}
&&\mathrm{exp}( 2\pi i \int_{\Sigma} ( q_{1} F + g G ) ) \nonumber\\
= &&\mathrm{exp}( 2\pi i \oint_{ \partial \Sigma} ( q_{1} A + g a )
) \prod_{U_{i} \cap U_{j} \cap U_{k} \cap \Sigma \ne 0} S(ijk).
\quad
\end{eqnarray}
The crucial problem is that the two factors within Eq. (19) can
become dependent on the choice of neighborhoods although their
choices are independent of the right hand side. It is argued that
when adding a neighborhood to the interior of $ \Sigma $, it cannot
give an effect on the second factor and cannot unambiguously become
affected to the first factor. But the more serious problem arises up
provided that we can take an addition of a neighborhood to the covering of the
boundaries $ \partial \Sigma_{i}, \forall i = 1, 2, \dots, L $. This
fact forces us to change the sign of both factors. Thus we cannot
deal with the two factors individually as the intrinsic properties
of given manifolds. In order to resolve this problem,
if we could have a square of Eq. (19), the issue of sign can be well defined due to
a double covering.

However, even if we cannot have any better idea for
understanding Stokes' theorem in the current context, there can very useful
if the boundary of $ \Sigma_{i} $ has a two cycle that can be represented as a double covering,
say, $ \partial \Sigma_{i} = 2 \gamma_{i}, \forall i = 1, 2, \dots, L
$. As a matter of fact, this boundary quantization scheme is mathematically
a mapping to lift the $ U(1) \times SU(N_{1}^{c}) $ gauge bundle
to a total covering space that combines the gauge bundle manifold into the basis one.
On the double covering condition, the boundary Dirac quantization gives rise to a form
\cite{alvarez,am}
\begin{eqnarray}
&&\mathrm{exp}( 2\pi i \int_{\Sigma} ( q_{1} F + g G ) ) \nonumber\\
&=&(-1)^{\omega(\Sigma)}\mathrm{exp}( 2\pi i \oint_{ \partial
\Sigma} 2( q_{1} A + g a ) ).
\end{eqnarray}
Under the fact that the two factors are independent of choices of
neighborhoods, they can be well defined since the second factor is
well-behaved in the case of inserting the number 2 in the exponent of equation (20).
It is claimed that $ \omega(\Sigma) $ becomes well-defined mod 2 on the $
\Sigma $ with the boundary quantization condition as a double covering map.

Now we apply the boundary quantization scheme of double covering to
the Dirac flux quantization of $ U(1) \times SU(N_{1}^{c}) $ gauge theory
on a spin manifold with 2-cocycle boundary, i.e.,
$ \partial \Sigma_{i} = 2 \gamma_{i}, \forall i = 1,
2, \dots, L $. Hence the flux quantization of double covering leads to a form
\cite{park2,alvarez,am}
\begin{eqnarray}
&&\mathrm{exp}( 2\pi i \int_{\Sigma_{i}} ( q_{1} F + g G ) ) \nonumber\\
&=&(-1)^{\omega(\Sigma_{i})}\mathrm{exp}( 2\pi i \oint_{ \gamma_{i}}
2( q_{1} A + g a ) ), \forall i = 1, 2, \dots, L \qquad
\end{eqnarray}
where $ \omega(\Sigma_{i}) $ is a second Stiefel-Whitney class of $ \Sigma_{i} $.
It follows that we can obtain $ q_{1} = \frac{e}{2 N_{1}^{c}} $ from the even,
i.e., 2-cycle, flux quantization at the boundary on a 4D manifold with spin structures.
This remarkable result reveals that the system can have degenerate
ground states on closed topologically nontrivial
space-time-reversal-protected gapless surface states that are
characterized by $ (-1)^{\omega(\Sigma)} $ at the boundary. In a TRS
TI, $ Z_{2} $ topological objects of fractionalized charge can
emerge from the 2-cocycle flux quantization at the boundary.

It has been known that an effective FT can have a very serious problem of TRS broken
spontaneously because of strongly interacting nonabelian $ SU (N_{1}^{c}) $ \cite{witten0}.
In order to resolve the broken TRS problem, we have focused on a more fundamental concept
and deep insight concering Dirac quantization conditions over a Riemannian manifold with
spin structures of topological property. These spin structures
can naturally provide the second Stiefel-Whitney characteristic classes for the conditions of
flux quantization in the non-empty boundary manifold with spin structures. Since
the gauge bundle $ SU (N_{1}^{c}) $ can naturally be connected to a basis manifold,
we have exploited a mathematical method to lift it to a total covering space that unifies
the gauge bundle manifold into the basis one. This lifting scheme can lead to the
quantization condition on a strongly correlated nonabelian gauge theory
without having the seriously broken TRS.
Therefore, under the quantization conditions of complex spinor
fields, we can construct an effective topological QFT for studying
a nonabelian gauge field theory of strongly correlated electrons on
the Riemannian manifold with nonempty boundary spin manifold in correlated TIs.

\section{\label{sec:level1} IV. Spin Chern-Simon Thoery and Index Theorem of Effective Field Theory }

Thus far we have investigated the Dirac flux quantization with the
total $ U(1) \times SU(N_{1}^{c}) $ quark field on the complex
spinor representations of antisymmetric parton wavefunctions.
The interactions yield the quarks to condense at low energies into a
noninteracting T-invariant TI state with axion angle $ \theta $.
Using Dirac flux quantization, index theorem, and spin CS
theory, i.e., half-integer CS theory, we manipulate an effective
topological QFT for the total $ U(1)
\times SU(N_{1}^{c}) $ quark field strength $ q_{1} F + g G $ on the
complex spinor representations of antisymmetric parton wavefunctions.
It is assumed that $ M _{4} $ has spin structures with a 2-cycle boundary of
double covering. In particular, we should include the above results of
the 2-cycle boundary quantization into the topological term of the FT.
In the topological term $ \frac{\theta}{2 \pi}
\frac{e^{2}}{2 \pi} \mathrm{E} \cdot \mathrm{B} = \frac{\theta
e^{2}}{32 \pi^{2}} \epsilon^{\mu \nu \rho \sigma} F_{\mu \nu}F_{\rho
\sigma} $ for noninteracting TIs, the $ U(1) $ electron field
strength is replaced by $ U(1) \times SU(N_{1}^{c}) $. Let us
construct a partition function
\begin{eqnarray}
\mathcal{Z} = C (-1)^{\omega} \mathrm{exp}(
i\int_{M_{4}} d^{3}x dt \mathcal{L}_{eff} (F,G) ).
\end{eqnarray}
An effective Lagrangian for the gauge theory of total quarks is
expressed by $ \mathcal{L}_{eff} = \mathcal{L}_{0} +
\mathcal{L}_{\mathrm{top}} $ where $\mathcal{L}_{0} =
-\frac{1}{4}G_{\mu \nu}G^{\mu \nu} $ is the kinetic Yang-Mills
Lagrangian. The second topological term of $ \mathcal{L}_{eff} $ has
a following form
\begin{eqnarray}
\mathcal{L}_{\mathrm{top}} &=& \frac{\theta_{1}}{32 \pi^{2}}
\epsilon^{\mu \nu \rho \sigma} \mathrm{Tr}[( q_{1} F_{\mu \nu} + g
G_{\mu \nu})( q_{1} F_{\rho \sigma} + g G_{\rho \sigma} )]
\nonumber\\
&=& \partial_{\mu} \epsilon^{\mu \nu \rho \sigma} \{
\frac{\theta_{eff} e^{2}}{8 \pi^{2}}
A_{\nu}\partial_{\rho}A_{\sigma} + \frac{\theta_{1} g^{2}}{8
\pi^{2}} (a_{\nu}\partial_{\rho}a_{\sigma} + \mathcal{L}_{\nu \rho
\sigma}) \}, \quad
\end{eqnarray}
where $ \mathrm{Tr} $ denotes the trace in the $ N_{1}^{c} $
representation of $ SU(N_{1}^{c}) $, $ F_{\mu \nu} = \partial_{\mu}
A_{\nu} -
\partial_{\nu} A_{\mu} $ and $ G_{\mu \nu} =\partial_{\mu} a_{\nu} -
\partial_{\nu} a_{\mu} + ig[a_{\mu},a_{\nu}] $ indicate $ U(1) $ and
$ SU(N_{1}^{c}) $ field strengths, respectively, and $ \theta_{1} $
is an action angle for $ N_{f}$ =1. Here we denote that $
\theta_{eff} = \frac{\theta_{1}}{2N_{1}^{c}} $ and $
\mathcal{L}_{\nu \rho \nu} = i \frac{2}{3} g
a_{\nu}a_{\rho}a_{\sigma} $. It is also noted that crossed terms
such as $ \mathrm{Tr}( F_{\mu \nu}f_{\rho \sigma}) $ vanishes owing
to the tracelessness of the $ SU(N_{1}^{c}) $ gauge field. Using
spin CS theory, we rewrite the partition function
\begin{eqnarray}
\mathcal{Z}
&=& C (-1)^{\omega} \mathrm{exp}[ i\int_{M_{4}} d^{3}x dt
\mathcal{L}_{0} + i \theta_{1} g^{2}
CS^{s}(a) ] \nonumber\\
&& \times \mathrm{exp} i\int_{M_{3}} \{ \frac{\theta_{eff} e^{2}}{8
\pi^{2}} A \wedge dA + \frac{\theta_{1} g^{2}}{8 \pi^{2}}
\mathcal{L} \},
\end{eqnarray}
where $ CS^{s}(a) $ is a spin CS action defined by $ \frac{1}{8
\pi^{2}} \int_{M_{3}} a \wedge da $ on a 3D manifold with spin
structure $ M_{3} $ which is a boundary of $ M_{4} $.

Concerning the spin CS action, Jenquin developed spin CS theory over
the space of connections on the basis of the theorem of Dai and
Freed regarding the $ \xi $-invariant of the Dirac operator
\cite{dai,jenquin}. Exploiting these results, the spin CS actions
and the $ \xi $-invariant can be identified up to multiplication
with a metric dependent function $ f(h_{a}) $ which is given by
\cite{belov}
\begin{eqnarray}
e^{2 \pi i CS^{s}(a)} = e^{i \pi \xi({D_{a}}) } f(h_{a}),
\end{eqnarray}
where $ D_{a} $ is a Dirac operator on a compact 3D manifold $ M_{3}
$, and $ h_{a} $ is a metric on it. The $ \xi $-invariant is
described by $ \xi(D_{a}) = \frac{1}{2} ( \eta_{D_{a}}(0) +
\mathrm{dim} \mathrm{ker} D_{a} ) $. Here $ \eta_{D_{a}}(0) $ refers
to the $ \eta $-invariant of $ D_{a}(0) $ defined by $
\eta_{D_{a}}(s) = \Sigma_{\tilde{\lambda} \ne 0} \frac{\mathrm{sign}
\tilde{\lambda} }{ |\tilde{\lambda}|^{s}}, \mathrm{Re}(s) >
\frac{3}{2} $. In 3 D, it follows that $ \xi(D_{a}) $ mod 2 is also
a smooth function of the geometric parameters.

In order to understand the singular behavior of $ \xi({D_{a}}) $,
let us consider a singular Dirac monopole in a strongly correlated
topological insulator. The singular Dirac is a classical solution of
Maxwell's equations that becomes invariant under the translations
and rotations of $ \vec{\mathrm{r}} $, and takes singularity at the
line defined by $ \vec{\mathrm{r}} = 0$. Assume that $ G_{U(1)} $ is
a curvature of a $ U(1) $ connection $ a $. Then, Maxwell's
equations can be solved on $ \mathbb{R}^{3} \diagdown {0} $, i.e.,
on the complement of the point $ \vec{\mathrm{r}} = 0$ in $
\mathbb{R}^{3} $, by $ G_{U(1)} = \frac{i}{2} *
d(\frac{1}{|\vec{\mathrm{r}}|})$ where $ * $ is a Hodge operator
that explicitly gives the dependence on the orientation of the
normal bundle to the line. If $ Y $ is a two-sphere enclosing the
singularity around $ \vec{\mathrm{r}} = 0$, then $ \frac{i }{2 \pi }
\int_{Y} G_{U(1)} = 1 $. The $ U(1) $ field allows us to consider
dyons that are bound states of electric and magnetic charges. In
particular, the Witten effect of TIs produces charge $
\frac{\theta}{2 \pi} \frac{e g_{mim}}{2 \pi} e $ to the monopole
where the minimal strength is written by $ g_{min} = \frac{4
\pi}{e}$ \cite{witten1}. The singular part of $ G_{U(1)} $ enables
us to take $ G_{U(1)} \thickapprox \frac{ine}{4}
* d(\frac{1}{|\vec{\mathrm{r}}|}) $ due to the $ \theta $ term of
the Witten effect. It follows that there can be two topological
objects with fractional charge $ ne/2 $ in the strongly correlated
TI. The TI ground state has a gap to all topological excitations,
and preserves the TRS. It should take ground state degeneracy (GSD)
on topologically non-trivial spaces with non-contractible loops
corresponding to arbitrary cycles. The ground states can be
represented by configurations $ (\{\gamma_{i}, n_{i}\}) $ of $ Z_{2}
$ flux through the non-contractible loops $ \gamma_{i}, \forall i =
1, 2, \dots, N $. Therefore, these topological objects can be
described by the representation of a holonomy $
\mathrm{Hol}_{a}(\{\gamma_{i},n_{i}\}) $ of $ a $ around any loop $
\gamma_{i} $. Furthermore the $ (-1)^{\omega} $ factor of
Eq. (24) can be replaced by this holonomy.

Accounting for the holonomy of the topological objects, and
substituting Eq. (25) into Eq. (24), the partition function yields
\cite{belov}
\begin{eqnarray}
\mathcal{Z} &=& C(-1)^{ \frac{ \xi(D_{a}) g^{2} \theta_{1} }{2\pi} }
\mathrm{Hol}_{ a(\{ \gamma_{i},n_{i} \})} \cdot K(h_{a},\theta,g) \nonumber\\
&&\times \mathrm{exp}( i\int_{M_{4}} d^{3}x dt \mathcal{L}_{0} +
i\frac{\theta_{eff} e^{2}}{8 \pi ^{2}} \int_{M_{3}} A \wedge dA ),
\quad
\end{eqnarray} where $ K(h_{a},\theta,g) $
denotes
\begin{displaymath}
 K(h_{a},\theta,g) =[f(h_{a}) \mathrm{exp}(
-\frac{2}{3} g \int_{M_{3}} a \wedge a \wedge a ) ]^{ \frac{
\theta_{1} \lambda_{1}^{c} }{ 2\pi N_{1}^{c} } },
\end{displaymath}
with $ \lambda_{1}^{c} \equiv N_{1}^{c} g^{2} $ fixed. As $
N_{1}^{c} \rightarrow \infty $ under the fixed $ \lambda_{1}^{c} $,
$ \frac{\theta_{1} \lambda_{1}^{c} }{ 2 \pi N_{1}^{c} } \rightarrow
0 $. Thus it is shown that $ K(h_{a},\theta,g) $ goes to 1. These
facts give rise to the partition function in a simple form
\begin{eqnarray}
\mathcal{Z} &&\approx C(-1)^{ \frac{ \xi(D_{a}) g^{2} \theta_{1}
}{2\pi} } \mathrm{Hol}_{ a }(\{ \gamma_{i},n_{i} \})
\nonumber\\
&& \times \mathrm{exp}( i\int_{M_{4}} d^{3}x dt \mathcal{L}_{0} + i
\frac{\theta_{eff} e^{2}}{8 \pi ^{2}} \int_{M_{3}} A \wedge dA ).
\end{eqnarray}

\section{\label{sec:level1} V. Fractional Magnetoelectric Effects and Fractional Surface Quantum Hall Effect }

One can study fractional magnetoelectric effects and surface quantum Hall effect
by means of the partition function described by Eq. (27) on the effective topological QFT in
strongly correlated TIs. The electromagnetic response allows one to calculate the effective
axion angles $ \theta_{eff} $. The T-invariance provides one for quantization of
$ \theta_{eff} $ in integer multiples of $ \pi $ provided
that the minimal electric charge becomes an effective one $ e_{eff} $
owing to the holonomy and under the minimum integer case of $ n_{i}
= 1$ in Eq. (27). For $ \theta_{1} $, one can make the following
choices using the current parton model of quarks in the absence of
breaking the TRS. It follows that there can be two kinds of choices
such that $ \frac{ \xi(D_{a}) g^{2} \theta_{1} }{2\pi} = 2k_{1} $ or
$ 2k_{1} + 1 $ for any $ k_{1} $. Thus for the choice of the lowest
value of $ k_{1} = 0$ in $ \theta_{1} $, the effective axion angles
are given by two expressions
\begin{eqnarray}
\theta_{eff} = \frac{\theta_{1}}{2 N_{1}^{c}} = 0, \quad  \frac{
2\pi }{ \xi(D_{a})  \lambda_{1}^{c} }.
\end{eqnarray}

Let us generalize the $ N_{1}^{c} $ case to the effective field
theory for multiple flavor values $ N_{f}^{c} \ge 1 $. Assume that
quarks of flavor $ f $ produces a noninteracting TI with $
\theta_{f} = \frac{ 2 \pi (2k_{f} +1)}{ \xi(D_{a}) g^{2} }, \forall
k_{f} $. After integrating them, the effective theory has a gauge
group of $ U(1) \times \prod_{f=1}^{N_{f}} U(N_{f}^{c})/U_{e}(1) $
where $ U(1)_{e} $ indicates the overall $ U(1) $ gauge
transformation of the electron operator. This gauge group yields the
electromagnetic axion angle $ \theta_{eff} = \{ \sum_{f=1}^{N_{f}}
\frac{N_{f}^{c}}{\theta_{f}} \}^{-1} $. If $ \theta_{f} $ is
considered to be $ \theta_{f} = \frac{ 2\pi}{ \xi(D_{a}) g^{2} } $
for each flavor, then there can be degenerate axion angles of even
integers such as $ \theta_{eff} = \frac{ 2\pi}{ \xi(D_{a}) }
\sum_{f=1}^{N_{f}} \frac {1}{2\lambda_{f}^{c}} $ where $
\lambda_{f}^{c} \equiv N_{f}^{c} g^{2} $. Thus, the effective axion
angle yields a simple expression
\begin{eqnarray}
 \theta_{eff} = \frac{ 2 \pi }{ \xi(D_{a}) (2g^{2} N(N_{f}, c))}
 \equiv \frac{ \pi }{ \xi(D_{a}) \lambda(N_{f},c) },
\end{eqnarray}
where $ N(N_{f}, c) $ indicates $ N(N_{f}, c) = N_{1}^{c} + \cdots +
N_{N_{f}}^{c} $ as an odd number, and $ \lambda(N_{f},c)
\equiv g^{2} N(N_{f}, c) $.

The partition function of the gauge fields enables one to determine
important physical properties for the interacting TI. In general,
the surface of the interacting TI has an action domain wall with the
electromagnetic action angle which jumps from $ \theta_{eff} $ in
the fractional TI to 0 in the vacuum. Hence by flux quantization,
the domain wall of interacting TI leads to the surface QHE
\begin{eqnarray}
\sigma_{H}^{s} = \frac{\theta_{eff}}{2 \pi} \frac{e^{2}}{h} =
\frac{1}{2 \xi(D_{a})\lambda(N_{f},c)} \frac{e^{2}}{h}.
\end{eqnarray}
If $ \xi(D_{a}) $ = 2 in 3D, then the surface QHE gives rise to a
simple form
\begin{eqnarray}
\sigma_{H}^{s} = \frac{1}{4 \lambda} \frac{e^{2}}{h}.
\end{eqnarray}
In the viewpoint of the magnetoelectric
response for the interacting TIs, the fractional axion angle allows
one to result in the fractional magnetoelectric polarization $
K_{\theta} = \frac{1}{4 \lambda} $. It is proposed that the
factional magnetoelectric effect can be realized in gapped time
reversal invariant topological insulators of strongly correlated
bosons or fermions with an effective axion angle
$ \theta_{eff} = \frac{\pi}{2\lambda}$ if they can
have fractional excitations and degenerate ground states on
topologically nontrivial and oriented spaces.

On the other hand, provided that an effective charge can be expressed in terms of
$ e_{eff} = \frac{e}{\sqrt{2 \lambda}} $, the surface Hall
conductance is regarded as a conductivity $ \sigma_{H}^{s} =
\frac{e_{eff}^{2}}{2h} $ for a single Dirac cone of gapless fermions
with fractional charge $ e_{eff} $ and a pure axion angle $ \theta = \pi $.
Recently, angular photoemission spectroscopy experiment has showed that
there is a topological phase transition from a trivial insulator into
a topological surface state by displaying the emergence of
spin vortex with fractional charge $ \pm \frac{e}{2} $  and $ \theta = \pi $
in a tunable topological insulator $ \mathrm{BiTl(S_{1-\delta}Se_{\delta})} $ \cite{xu}.
Theoretically it has been suggested that there can be a quantized vortex of fractional charge
$ \pm \frac{e}{2} $ and an odd number of gapless Dirac fermions at the surface
of a strong TI \cite{seradjeh}. In the viewpoint of these experimental
and theoretical studies, if $  \lambda(N_{f},c) = 2 $ as well as $ \xi(D_{a}) $ = 2, then one
can obtain a well-known formula for the surface Hall conductivity of
massless Dirac fermions
\begin{eqnarray}
\sigma_{H}^{s} = \frac{\tilde{e}^{2}}{2h},
\end{eqnarray}
where $ \tilde{e} = \frac{e}{2} $ as a fractional charge.
This theoretical construction shows
that the TIs can have fractionalized charge $ \frac{e}{2} $ topological
objects with a bulk gap and string-like vortex excitations.
These topological excitations can be described in terms
of a deconfined $ Z_{2} $ gauge theory in (3+1)D.
Due to topologically protected gapless surface states, the unexpected
result above has been emerged from the bulk $ \theta $ term on a new
topological phase of quantum matter in (3 + 1)D. If the surface of
strongly correlated TIs is covered by time-reversal breaking
materials, the bulk $ \theta $ term enables us to observe the
surface Hall conductivity of Eq. (32).

So far, in a fundamental view point, we have focused a topological invariant
on the boundary of the strongly correlated TIs. The topological invariance
provides us for some interesting physical properties without
knowing the microscopic Hamiltonian of the strongly correlated systems.
In this article, one of the main focuses is to study the topological invariance
on the boundary of the strongly correlated TIs with nonabelian gauge theories
without taking the microscopic Hamiltonian of TIs in details.
Under the topological invariance, the effective action can be written
as a full function of $ g^{2}N(N_{f},c) $ \cite{maldacena,mcgreevy}
\begin{eqnarray}
S_{eff} = -\sum_{\mathrm{g}=0}^{\infty}N(N_{f},c)^{2-2\mathrm{g}}
F_{\mathrm{g}}(g^{2}N(N_{f},c))
\end{eqnarray}
where the sum over topologies becomes explicit.
In Eq. (33), the 't Hooft limit gets to be $ N(N_{f},c) \rightarrow \infty $
and a 't Hooft coupling $ \lambda = g^{2}N(N_{f},c)$, fixed \cite{maldacena,mcgreevy}.
$ F_{\mathrm{g}}(g^{2}N(N_{f},c)) $ is a scaling function of $ \lambda $.
Thus we can represent the effective action of $ g^{2}N(N_{f},c) $ 
as the fixed value of $ \lambda $, and as a sum over surface topology.

Let us consider a surface Hall conductance on an oriented surface withe genus 
in topology. Then the key idea of topological consideration is to investigate
the surface Hall conductance of topological objects on the surface of 
the genus $ \mathrm{g} $ in correlated TIs with spin structures. 
There can be geometric cycles for all even codimensions in 
general complex structures on a given surface. Todd defined polynomials 
in theses cycles which can represent for a variety of dimension  
a certain number called the Todd genus \cite{todd}. The Todd genus can be 
expressed in terms of the Chern classes of tangent bundles on the fundamental 
cycle as the Euler characteristic of a differentiable manifold.

To measure the GSD on topologically nontrivial spatial 3-manifolds,
consider a fractional TI on a manifold $ \Sigma_{\mathrm{g}} \times
I $ where $ \Sigma_{\mathrm{g}} $ is a Riemann surface of Todd genus $
\mathrm{g} $ and $ I = [0,t] $ stands for a bounded interval when $
t $ is the sample thickness, and two copies of $ \Sigma_{\mathrm{g}}
$ are two bounding surfaces at each ends of $ I $. A noninteracting
TI with a $ \nu = \frac{1}{2} $ Laughlin state deposited on both
surfaces is described by two independent CS theories \cite{belov}
and has a ground state of $ 2^{\mathrm{g}} $ on each surface for a
total GSD of $ (2^{\mathrm{g}})^{2} = 2^{2\mathrm{g}} $ under the Todd genus. 
This is a disjoint union of the tori which is called the Picard group. 
In the $ \mathrm{g}-1 $ component of the Picard group, the spin line
bundles are represented by $ 2^{2\mathrm{g}} $ points. Due to the
Atiyah-Patodi-Singer index theorem \cite{aps}, the variation of the
$ \xi $-invariant under the change of smooth parameters gives rise
to $ \xi(D_{a},t) - \xi(D_{a},0) = \int_{ M_{3}\times [0,t] }[
\hat{A}( \frac{R(s)}{2 \pi} ) \wedge \mathrm{ch}( G_{a}(s) ) ] $
where $ \hat{A} $ is the roof-genus as the Todd class of an almost 
complex manifold. $ R(s) $ is the Riemann tensor
corresponding to the metric $ h(s) $ and $ \mathrm{ch}( G_{a}(s) ) $
is a Chern class \cite{belov}. Therefore it is shown that there can
exist a genuine fractional TI such that $ K_{\theta} = \frac{1}{
2\lambda(N_{f},c) (\xi(D_{a},t) - \xi(D_{a},0))} $ and Hall
conductance
\begin{eqnarray}
\sigma_{H}^{s} = \frac{1}{2\lambda(N_{f},c) (\xi(D_{a},t) -
\xi(D_{a},0)) } \frac{e^{2}}{h}.
\end{eqnarray}

\section{\label{sec:level1} VI. Summary and Conclusion }

Having exploited flux quantization, index theorem and spin CS theory,
we construct an effective topological
field theory of strongly correlated topological insulators coupling
to a nonabelian gauge field $ SU(N) $ with an interaction constant $
g $ in the absence of the TRS breaking. If $ N $ and $ g $ allow us
to define a t'Hooft parameter $ \lambda $ as $ \lambda = N g^{2} $,
then our construction leads to FQHE on the surface with Hall
conductance $ \sigma_{H}^{s} = \frac{1}{4\lambda} \frac{e^{2}}{2h}
$. In particular, if $ \lambda =2 $, then it has been shown that $
\sigma_{H}^{s} = \frac{\tilde{e}^{2}}{2h} $. For the magnetoelectric
response described by a bulk axion angle $ \theta $, it has been
proposed that the factional magnetoelectric effect can be realized
in gapped time reversal invariant topological insulators of strongly
correlated bosons or fermions with $ \theta = \frac{\pi}{2\lambda}$
if they can have fractional excitations and degenerate ground states
on topologically nontrivial and oriented spaces.
Thus there can exist
degenerate ground states of FQHE on the surface with Hall
conductance of Eqs. (31),(32) and (34). Under our current effective
field theory, it is claimed that the topological band structure from
the nontrivial Hamiltonian matrix may provide the nontrivial
properties of the quantum states with fractionalization and the
emergent nonabelian gauge theory without the microscopic models of
strongly correlated TIs. As one of future works, it would be very
interesting to apply the current theory to the
confinement-deconfinement issue in topological quantum
chromodynamics by the $ SU(N) $ nonabelian gauge field.

This work was supported by the Basic Science Research Program (2009-0083512),
by Priority Research Centers Program (2010-0029711) through the National Research Foundation
of Korea (NRF) funded by the Ministry of Education, Science and Technology,
and by the Brain Korea 21 Project in 2011.


\end{document}